\documentclass[10pt,journal,compsoc]{IEEEtran}

\usepackage{subfiles,cite}
\usepackage{graphicx}
\usepackage{amsthm}
\usepackage{booktabs} 
\usepackage{graphicx, epsfig, color, amsmath}
\usepackage[ruled,vlined]{algorithm2e}
\usepackage{caption}
\usepackage{subcaption}
\usepackage{hyperref}

\theoremstyle{definition}
\newtheorem{definition}{Definition}[section]
\theoremstyle{definition}
\newtheorem{example}{Example}[section]

\SetCommentSty{mycommfont}

\title{\Large \bf
``Do You Know You Are Tracked by Photos That You Didn't Take": \\Location-Aware Multi-Party Image Privacy Protection
}

\author{\IEEEauthorblockN{Joshua Morris$^\dag$, Sara Newman$^\dag$, Kannappan Palaniappan$^\dag$, Jianping Fan$^\ddag$, Dan Lin$^\dag$}\\
\IEEEauthorblockA{$^\dag$ Department of Electrical Engineering and Computer Science \\
University of Missouri \\
{\em jdm6b3@mail.missouri.edu,sn976@mst.edu,pal@missouri.edu,lindan@missouri.edu} \\
$^\ddag$Department of Computer Science\\
     University of North Carolina - Charlotte\\
{\em jfan@uncc.edu}}
}

\begin{document}

\maketitle
\thispagestyle{empty}
\pagestyle{empty}

\begin{abstract}

Most existing image privacy protection works focus mainly on the privacy of photo owners and their friends, but lack the consideration of other people who are in the background of the photos and the related location privacy issues. In fact, when a person is in the background of someone else's photos, he/she may be unintentionally exposed to the public when the photo owner shares the photo online. Not only a single visited place could be exposed, attackers may also be able to piece together a person's  travel route  from images. In this paper, we propose a novel image privacy protection system, called LAMP, which aims to light up the location awareness for people during online image sharing. The LAMP system is based on a newly designed location-aware multi-party image access control model. The LAMP system will automatically detect the user's occurrences on photos regardless the user is the photo owner or not. Once a user is identified and the location of the photo is deemed sensitive according to the user's privacy policy, the LAMP system will intelligently replace the user's face. A prototype of the system was implemented and evaluated to demonstrate its applicability in the real world.

\end{abstract}

\section{Introduction}

With the growing ubiquity of smartphones and other mobile devices, image sharing  is gaining increasing popularity in social networks like Facebook, Instagram and Foursquare. During social image sharing, privacy protection has now become a crucial issue to be addressed since images can intuitively tell {\em when} and {\em where} a special moment took place, {\em who} participated and {\em what} were their relationships, i.e., sharing images can reveal much of users' personal and social environments and their private lives \cite{besmer-facebook-2008}. News has reported multiple incidents about people being fired due to their private photos being disclosed to undesired audience \cite{fired-social-media,teacher-bikini}.


Recognizing the importance of image privacy, researchers and social media sites have developed various privacy policies and tools to help users specify the group of people for photo sharing. However, most existing image privacy protection approaches \cite{context-dependent-ml, a3p, privacy-level,user-defined-sharing,cenlocshare,sharing-google+,hideme,iprivacy,multiparty-authorization-2011,multiparty-access-2013,face-off} focus mainly on the privacy of photo owners and at most the photo owners' friends. They lack the consideration of other people who are in the background of the photos and are not related to the photo owners. In fact, when a person is  in the background of someone else's photo, he/she may be unintentionally exposed to the public when the photo owner shares the photo online. For example,  Alice had a bad day and  visited a pub at night. Someone took a photo of the pub with Alice in the background. Alice had no idea about the photo until her supervisor came to show concerns to her because he coincidentally saw her drunk photo online posted by the other person. A recent interview \cite{next-meme} among college students also  confirmed such privacy concerns, indicating that more and more undergraduates worry about becoming an Internet meme because their embarrassing moments were photographed by their peers and posted on social media.  As an initial effort towards this new privacy problem,  Llia et al.  \cite{face-off} suggested the use of face recognition to identify all the people in the photo but their implementation is still limited to identifying photo owner's friends through available image tags and they have not considered the associated location privacy issues as discussed in the following.

With more and more images associated with geo-tags and timestamps, image privacy now comes to the crossroads of the location privacy. Such exposure may cause undesired consequences especially when the person being exposed was visiting sensitive locations. For example, a businessman Bob is meeting an important customer in a restaurant during a business trip while Jack, who usually reviews every restaurant he visits, took a photo of the restaurant with Bob and his customer in the background. Jack published his review along with  the photo to a social media site. Bob's company's competitor noticed Bob was in the photo. The photo may have leaked business intelligence since it  tells the competitor when and where Bob met the potential business partner whom the two companies are currently competing. There are many other scenarios, such as visiting a specialty hospital or attending alcoholic counseling,  which could cause similar uneasiness to the person if his/her photos at those sensitive locations were posted online by others.  Furthermore, attackers may even be able to piece together a person's  travel route by analyzing unprotected online photos. Specifically, the photos containing target victim's face may be identified via  face recognition; and the photos locations and timestamps may be revealed through various means such as geo-tags, metadata, or landmarks obtained from the advanced image processing tools. In Section \ref{sec:imagesharing}, we demonstrate an example of such an attack to show its feasibility.

To  better understand the aforementioned location related image privacy issues, we have conducted an exploratory user study among more than one hundred people to obtain their  privacy opinions over a set of scenarios.  The findings from the user study conform with our hypothesis that location sensitive photos could disclose too much of a person's privacy. Unfortunately,  there have been very little works on how to help users mitigate such location-dependent image privacy.  Thus, we  propose a novel image privacy protection system, called LAMP (Location-Aware Multi-party Privacy),  which aims to light up the location awareness for people during online image sharing. The LAMP system is based on a newly designed Location-Aware Multi-Party image (LAMPi) access control model that allows  individual user to specify sensitive locations and timestamps  for any photo in which their faces are identifiable. The proposed access control model goes beyond the traditional owner-centric privacy protection model, and the proposed LAMP system will facilitate social network providers to provide an equal protection for any people in the same photo.  Specifically, the LAMP system as an add-on to existing social media sites will automatically detect the user's occurrences on photos to be posted online regardless the user is the photo owner or not.  Once a user is identified and the location of the photo is deemed sensitive according to the user's privacy policy, the user's face will be replaced with a virtually generated human face. As we know, face blurring has been commonly used for privacy protection during photo sharing, while face replacement has been provided by existing apps mainly as a fun pastime activity. We hereby argue that the face replacement would be a better way to protect people's privacy as it offers several advantages which cannot be achieved by   face blurring.  First, it prevents attackers from using the latest image deblurring techniques  \cite{rank-deblurring,phase-deblurring,discriminative-deblurring} to uncover the people being protected. Second, the use of face replacement maintains the beauty and intact of the photo and reduces the chance of the photo to become a target of an attack. Considering that a photo with a blurred face and a photo with a swapped face, it is obvious that a blurred face has privacy concerns whereas a nicely swapped face may not even be noticed by the attacker.

The key challenge during the design of the LAMP system is how to achieve the location-aware privacy protection without affecting users' image sharing experience. First, we need to be able to obtain users' location privacy concerns without adding too much burden to users. For this, we design a graphic-based policy specification tool for users to easily specify sensitive locations at different granularity levels following our proposed LAMPi access control model.  Second, we need to ensure that  the image uploading time is not significantly delayed due to the privacy policy checking. To protect privacy for each person on the photo to be uploaded, the first step is to identify each person. However, given the huge number of social network users (e.g., 2.4 billion Facebook users), identifying a person who is not related to the photo owner by comparing the photo against all the social network users would be very time consuming and negatively impact the user's photo sharing experience. In order to overcome this problem, we propose a quick filtering approach that leverages face encoding and location policy indexing to drastically reduce the face comparison to a small group of candidates. We have implemented a prototype of the proposed system, and conducted a second user study. Our experimental results demonstrate the efficiency and effectiveness of our approach. In a summary, the contributions of our work are the following:
\begin{itemize}  \itemsep=0pt

\item We define a novel fine-grained location-aware multi-party image access control mechanism which breaks the existing limits of privacy protection only for photo owners and their friends by providing equal privacy protection to every identifiable individual in the photo instead of photo owners and their friends. Moreover, we consider the location-dependent privacy issues that are not studied in the past.

\item We build a proof-of-concept application, the LAMP, to automate the location-aware multi-party privacy protection process. The algorithms designed for LAMP are tested to be efficient and scalable to deal with the huge number of photos and users on social media sites.


\item We conducted two rounds of user studies involving more than 200 people to obtain valuable user opinions on location-dependent privacy issues and evaluate the effectiveness of privacy protection offered by our approach.

\end{itemize}

The rest of the paper is organized as follows. Section \ref{sec:imagesharing} presents the privacy risk analysis. Section \ref{sec:accesscontrol}  introduces  our proposed LAMPi access control model and its implementation. Section \ref{sec:database} describes the LAMP system. Section \ref{sec:privacy evaluation} reviews privacy evaluations. Section \ref{sec:efficiencyeval} reports experimental studies. Section \ref{sec:related} discusses related work.  Finally, Section \ref{sec:conclusion} concludes the paper.

\section{Image Sharing Risk Analysis}\label{sec:imagesharing}
\subsection{Threat Model}

As the saying goes, a picture is worth a thousand words. An online photo/image can give out rich information about {\em who} are doing {\em what} at {\em when} and {\em where}. To better analyze the privacy risks incurred by image sharing,  we classify image privacy based on two criteria: human-oriented, and context-oriented.

\vspace{5pt}
\noindent {\bf Human-oriented image privacy} can be further classified into three types:

  {\bf (1) Photo owner's privacy}:  This type of privacy is currently preserved by allowing the photo owner to specify the groups of people who are permitted to access the shared photo. Most of the research works  \cite{a3p,context-dependent-ml,privacy-level,multiparty-access-2013} and commercial social media sites provide policy recommendation and configuration tools to achieve this. For example, Facebook users can choose to share the photos only with their friends but not friends of friends.

 {\bf (2) Photo owner's friends' privacy}: This refers to the privacy of the photo owner's friends who took the photo together with the photo owner. For example, Alice plans to post a party photo that includes her friend Kate. Kate is a shy girl who rarely shares photos online. Considering Kate's privacy,  Alice may need to communicate with her before publishing the photo. However, such multi-party privacy issues are mainly discussed in academic world \cite{besmer-untagging-2010,multiparty-access-2013,multiparty-authorization-2011}.   The current social media sites offer very little functionalities that support the multi-party privacy protection.

{\bf (3) Unaware people's privacy}: This refers to the privacy of the people who are in the photo but are not aware of their photo being taken by others.  For example, when someone took a selfie on the street, other pedestrians may be captured in the photo. These pedestrians will not know when and where their photos would appear on the Internet. Recently,  an  interview-based study \cite{next-meme} among college students found  that undergraduates felt a heightened state of being surveilled by their peers when their photos were taken without their permissions and shared on social media by others. Participants in that study stated that they worried about being judged  by others in a negative way based on the images which they were not aware of being taken.

\vspace{5pt}
\noindent
{\bf Context-oriented image privacy} can be further divided  into two categories:

{\bf (1) Activity-dependant privacy}:  There are various scenarios when a person  does not feel comfortable of sharing that moment with everyone. For example, a person in a funny costume may just want to share the photo with his/her close friends. In another case, a girl was drunk and someone else took her photo \cite{next-meme}. If the photo was posted online, it could lead to misjudgement of the girl and damage her general reputation. News also reported that some people were fired due to online photos. One case is that a fireman took a sick day off for attending an event, and he was later fired because his supervisor saw the event photo and identified him \cite{fireman}.

 {\bf (2) Location-dependant privacy}: A photo can leak location information of a person in many ways. The photo's embedded EXIF (Exchangeable image file format) \cite{exifphoto}
is a  direct source that tells the date and GPS coordinates a photo was taken. Although some social media sites like Facebook and Instagram stripped the metadata when publishing the photos, they  store the metadata in
a separate database. If a hacker gains the access to these databases, it is even easier for them to track users since they now just need to look at the collection of metadata from all photos without spending much time on extracting metadata or analyzing photos one by one. Besides metadata, the  photo itself may tell where the location is. Advanced image processing algorithms can identify the landmarks and the street signs. Yet another way could be the crowd sourcing. People living in the neighborhood of the place where the photo was taken may easily spot familiar buildings on the photo. With this said, {\em posting photos without metadata is still not sufficient to guarantee the location-dependant privacy of people in the photo.}

Most existing works on image privacy mainly focus on protecting photo owner and their friends' privacy (detailed review can be found in Section \ref{sec:related}). Very limited efforts have been devoted into the other equally important privacy issues. i.e., unaware people's privacy, context-oriented image privacy. Thus, in this work, we aim to design a new access control mechanism to protect people's privacy in the photos which were taken without their knowledge and permissions. Our approach is complementary to existing methods and aims to achieve a full spectrum of image privacy protection. To better motivate our work, we will first present a location tracking attack and an exploratory user study in the following.

\subsection{Location Tracking Attack Using Photos}

\begin{figure}[!b]
    \includegraphics[width=.48\textwidth]{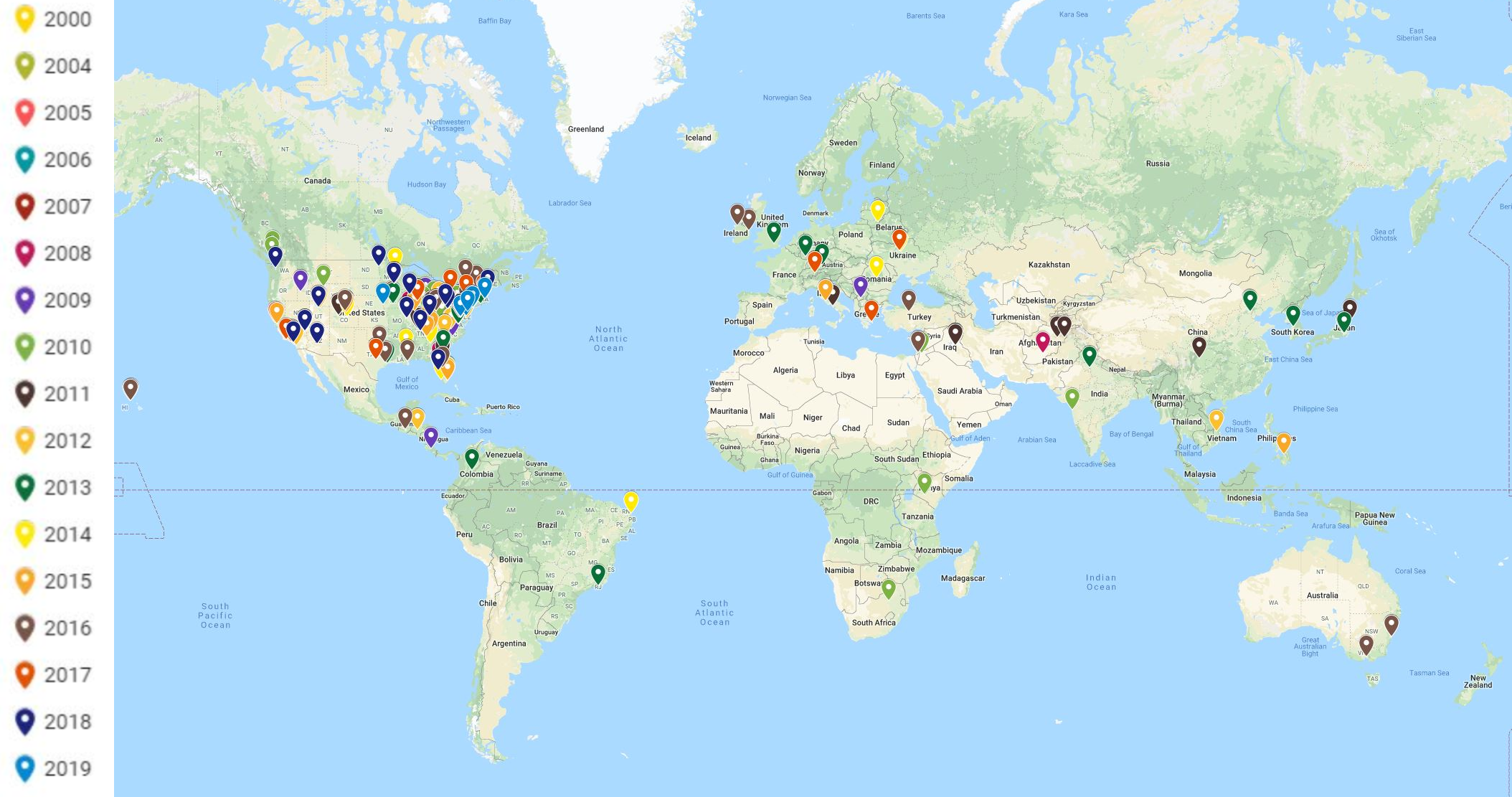}
    \caption{User Tracking Through Photo Metadata}
    \label{fig:metadata}
\end{figure}
In this experiment, we attempt to track a target person through his/her online photos. The goal is to show that it is not a difficult task for an attacker to cyberstalk a person. To avoid legal issues with a randomly chosen normal person, we decided to select a prominent figure whose images are publicly available in different venues: Joe Biden. Also, we do not hack into any social site to obtain metadata, whereas  attackers can certainly gain more information than us by doing so.

We wrote a script to automatically crawl Google images to obtain the target person's photos between 2000 and 2019. We collected around 30,000 photos for the target. From the collected  photos, we further analyze the metadata. Although not all the photos contain the metadata, it is still amazing that we were able to found 721 days of location and visiting time for the target. Based on the obtained information, we created a tracking map as shown in Figure \ref{fig:metadata}, where  each point on the figure shows the location of the target person and the color of the point indicates when the photo was taken.

It is worth noting that Google images may not return photos of a person if he/she is in the background. Even so, the photos obtained from Google images already reveal lots of location information of a person. When an attacker utilizes advanced image processing tools to look for any occurrences of a target (either foreground or background) and combines the knowledge of the metadata stolen from the social media providers, the target movement may be exposed in a much deeper level due to the prevalence of photographing nowadays. Considering that people who took photos of themselves know about the sensitivity of their current locations, whereas people who were in the background of others' photos have no idea their locations have been recorded, one idea in our proposed work is to  replace the faces of people who are in the background of the photos to avoid undesired exposure. In this way, even if the attackers run the image processing tool and have all the metadata of photos, the targets' faces have already changed and would not be identifiable.

\subsection{A User Study on Unexpected Privacy Disclosure}

In order to better understand users' concerns on location-dependent image privacy and gauge their interests in our proposed privacy protection mechanism, we conducted an online user study on Mechanical Turk. The user study is fully anonymous and follows the IRB exempted project guidelines.

We have recruited total 111 participants, including 51 females and 55 males. 15\% of them are between 18-25 years old, 41\% are between 26-35 years, 23\% between 36-45 years,  13\% between 46-55, and 8\% above 56. The age distribution conforms with the age groups of people who access the social media more often.


At the beginning of the user study, we asked participants if they were aware that online images may tell others where they were and what they were doing, and how much they valued their privacy especially location privacy.   From the response, we find that more than 74\% of participants were aware of the privacy issues incurred by online images, and more than 76\% emphasized that location privacy is important.

From there, we presented 10 different scenarios to the participants and asked them if they would be concerned when their images and their locations are disclosed to unexpected parties. Specifically, each scenario is accompanied with a short paragraph of story and an image. The 10 scenarios were designed with the goal to cover various aspects of our daily life in a nutshell.  From the participants' responses, we found that when a photo  discloses a critical moment or location of a person without being noticed by the person, more people would hope their identities are protected during the sharing of such kinds of  photos. For example, when someone (say Bob) is planning to switch a job and had a job interview at a restaurant, another customer who also dined at the restaurant took a photo of the restaurant with Bob and his interviewers in the background. The customer later shared his photo along with the restaurant review comments in a famous restaurant review website. If Bob's supervisor or colleagues saw the photo and recognized Bob and competitor company's people, that could raise  unnecessary tension in Bob's current workplace. Therefore, we see that 92.5\% of participants would like their identities be protected in this scenario, which is the scenario with the most concerns. The second mostly concerned privacy breach scenario is  when someone's children and home location may be exposed to strangers. About 91.5\% of participants desire an identity protection in this case. On the other hand, some scenarios that may not lead to severe consequences, such as having a personal trip and drinking at a bar, have received privacy concerns from a little fewer people, but still close to 70\%. Overall, we can observe that majority of people are concerned when their photos being taken and published by others without their knowledge, especially those photos that disclose sensitive locations and  reveal their private issues.

At the end of the survey, we also asked the following general question: ``{\em Suppose you are depicted in a photo published on social media by a stranger while you are in a  location where you wish not to be seen. If social media websites provided functionality for hiding your identity (e.g., face swapping) when you are  in such photos, would you like to use this function?"} More than 93.4\% of the participants  said that they would like to use such kind of services, which indicates a promisingly high acceptance rate of our proposed system.

\begin{figure*}[!t]
\centering
\includegraphics[width=6in]{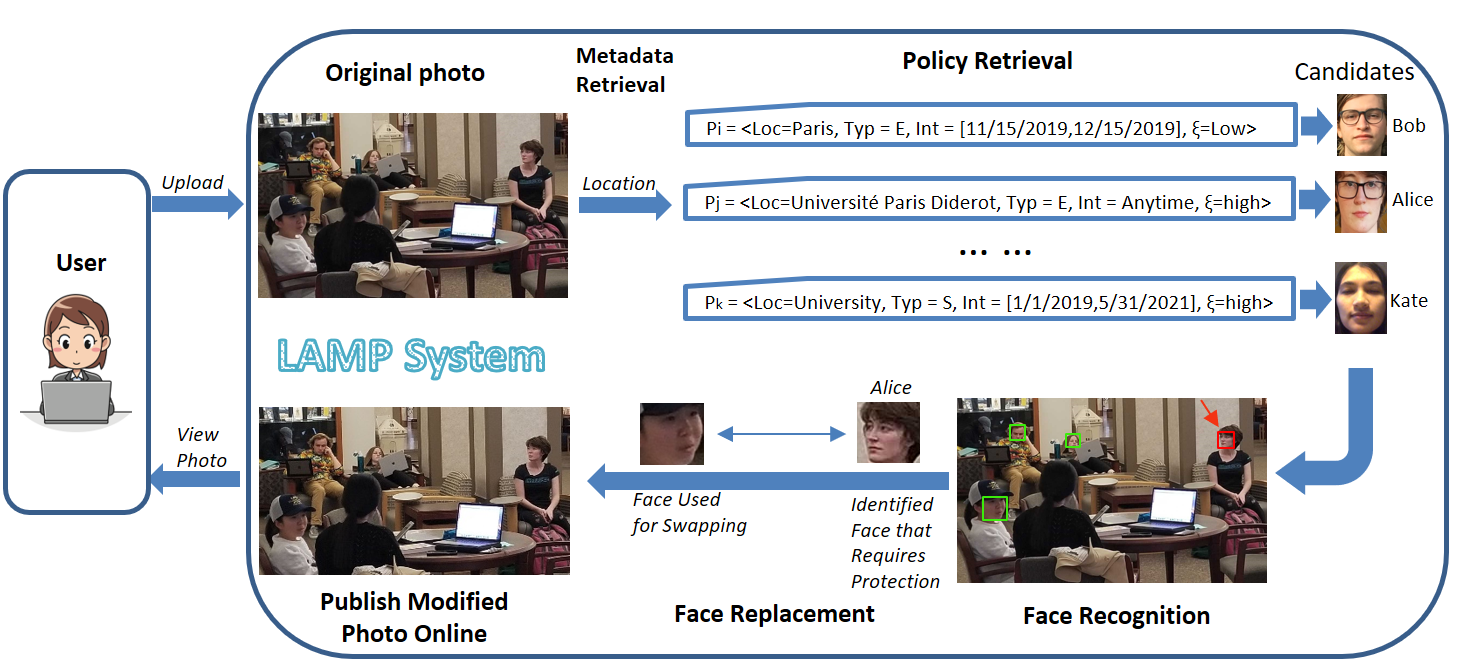}
\caption{The Privacy Protection Procedure in the LAMP System}
\label{fig:overview}
\end{figure*}

\section{LAMPi Access Control Mechanism}\label{sec:accesscontrol}

In the previous section, we have discussed both human-oriented and context-oriented image privacy, among which context-oriented image privacy of unaware people is least protected in the literature. To fill the gap,  we define the  LAMPi (Location-Aware Multi-Party image) access control mechanism, complementing  the traditional image access control. The LAMPi policies will allow users to specify location sensitivity at different scenarios and at different granularity levels. Its formal description is given below.
\begin{definition} A LAMPi policy $P$ of a user $u$ consists of the following components:

\noindent $\bullet$ Location range ($Loc$): The range of  locations protected by policy $P$.

\noindent $\bullet$  Location type ($Typ$): This indicates whether the location is given as a semantic location (denoted as `S')  or an exact address (denoted as `E').

\noindent $\bullet$  Time and date interval ($Int$): The time and date intervals during which the locations within $Loc$ should be protected.

\noindent $\bullet$ Sensitiveness ($\xi$): The sensitiveness of the location $Loc$, which has two levels:  ``High" or ``Low".
\label{def:policy}
\end{definition}

Since the LAMPi policies are mainly used to express the users'  privacy concerns when they are unintentionally captured in others' photos, there is no need for the LAMPi policy to specify the sharing group like traditional image privacy policies. Instead, the user can specify how much they care about themselves being exposed in the particular location using the sensitiveness level. When the sensitiveness level is set to high, the user's face at that location will be replaced for protection even if the user's face  on the photo is less identifiable, i.e., the face matching score is lower than a threshold (say 50\%). When the sensitiveness level is low, our system will only replace the user's face when the face matching score is above the threshold. In this way, we minimize unnecessary image modifications. For locations which are not specified in the user's policies, the locations are simply considered not sensitive for that user.


The following are some example LAMPi policies which demonstrate the usage of different policy settings.

\begin{example}{\em
Kate does not want others to post photos which show her  doing workout and sweating in a gym near her house. She can set her LAMPi policy as $P_{Kate}$=$\langle$Loc=\{gym\_address\}, Typ=E, Int=anytime, $\xi$=Low$\rangle$, where `Loc' is set to the gym's address, `Typ=E' indicates that this is an exact location, `Int=anytime' means Kate wants the privacy protection whenever she is in this gym, `$\xi$=low' means that  as long as she is not easily recognizable in the photo, Kate does not care about the photo being posted.  }

\end{example}

\begin{example}{\em
Alice does not wish her face being recognized on any online photo that shows  she is visiting a pub. She can thus set her LAMPi policy as $P_{Alice}=\langle$Loc=\{pub\}, Typ=S, Int=\{8pm-5pm on any day\}, $\xi$=High$\rangle$. That means if anyone attempts to post a pub photo with Alice in the background to a social network site, the social network site where Alice has registered and set the LAMPi policy will  automatically replace Alice's face with a synthetic face to preserve Alice's privacy without affecting the photo owner's sharing experience.  }
\end{example}



\section{The LAMP System}\label{sec:database}

In this section, we discuss how our proposed LAMP system helps preserve privacy of users who have no knowledge of their photos being posted by others. Figure~\ref{fig:overview} illustrates the data flow in  the LAMP system.  The LAMPi policy configuration function facilitates the users to specify the LAMPi policy through a graphic-based interface developed using Google Maps API. Users' policies will be indexed and stored in a policy database, and users' face features will be encoded to speed up the future face recognition. When someone wants to upload a photo to share,  our LAMP system will first retrieve policies which mark the photo location as sensitive. Among the owners of the retrieved policies, we will further check if their faces depicted on the photo. If so, their faces will be replaced with synthetic faces (or faces having no privacy concerns) to avoid undesired disclosure while maintaining the photo quality. In what follows, we elaborate the user identification and protection algorithms.


%


\begin{figure}[!t]
\centering
\includegraphics[width=3.4in]{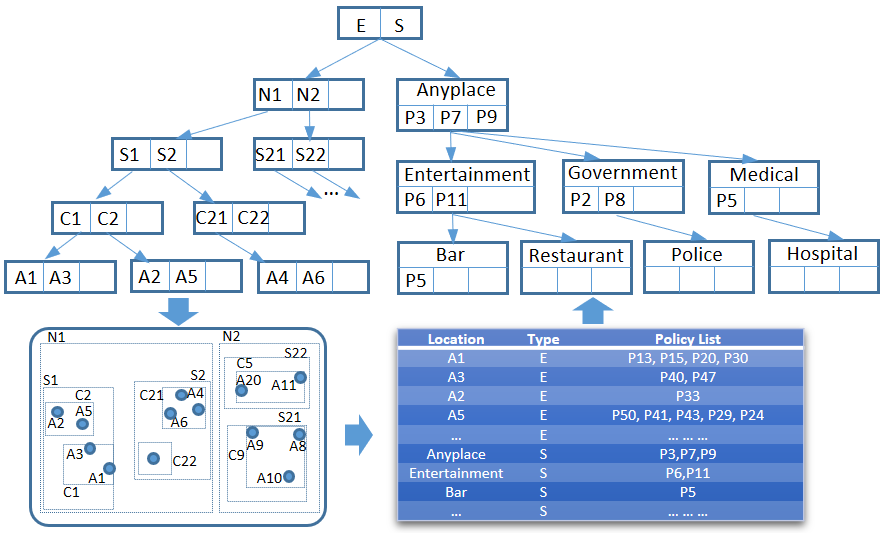}
\caption{An Overview  of the  DLP-tree}
\label{fig:dlp}
\end{figure}

\subsection{Identify Users  with Location Privacy Concerns}

In order to provide equal privacy protection to every person on the photo, an inevitable step is to know who (especially those in the background) are in the photo. A brute force method to identifying people in the background of the photo will need to compare the person's face on the photo against all the other users' faces in the social network site (e.g., 2.4 billion users in Facebook), which will be extremely computationally expensive and hardly possible to maintain real-time response for the photo uploading request. This has been a very challenging problem in the image privacy protection as also pointed out by Ilia et al. in \cite{face-off}.

To overcome this challenge, we aim to  reduce the total number of faces that need to be compared for each uploaded photo. Correspondingly, we propose a hybrid data structure that indexes policies and helps significantly reduce the search space,  making the face identification in large-scale user sets a feasible process for real-time applications.

\subsubsection{Indexing LAMPi Policies}

Given a photo and its location, we aim to quickly locate users who specify this location (based on the address) or this type of location (based on the semantic keywords) as sensitive so that later on we only need to compare these users' faces with those in the photo. To achieve this, we propose to index LAMPi policies based on locations, and group policies containing the same location together.  Specifically, we store the LAMPi policies in  a policy database implemented using PostgreSQL, and propose a DLP (Dual-Location-Policy) -tree to speed up the policy retrieval.

PostgreSQL was chosen for a variety of reasons. Most importantly is that PostgreSQL has great read and write performance compared to MySQL. Additionally, the PostGIS extension allows for geospacial data support that is incredibly useful when checking user and image locations. Finally, PostgreSQL easily supports concurrency in reading and writing that is crucial to speed up the face recognition process in our system as discussed in the later part of our approach.

The structure of the DLP-tree is illustrated in Figure \ref{fig:dlp}. The DLP-tree consists of two main parts to index exact locations and semantic locations in the LAMPi policies, respectively. The left side of the DLP-tree organizes exact locations in a hierarchical way from nation (N), state (S), city (C) to address (A).  Each entry in the leaf node is in the form of $\langle$$street$, $city$, $state$, $nation$, $\Gamma$, $PID$$\rangle$, where  the first four attributes are the address specified in the policy $PID$, and $\Gamma$ is the time and date interval when the location is considered sensitive. Nearby locations  are grouped together in the same leaf node or sibling leaf nodes. An entry in an internal node is in the form of $\langle$$region$, $CPT$$\rangle$, where $region$ described the region that covers all its child node pointed by $CPT$. In this way, the internal nodes can efficiently facilitate the LAMPi policy search.

Specifically, given a photo's location, we start the search from the root of the DLP-tree and traverse to the left side to find the node whose region encloses the photo's location. Then, we follow its child pointer to conduct the same boundary check in its child node until we reach the leaf level. In the leaf node, we further compare the photo's location against the locations specified in the LAMPi policies and retrieve those policies which mark the photo's location as sensitive. Note that the DLP-tree is different from a map since the DLP does not need to store all the places (e.g., all the addresses, all the cities) if no policies have been specified there yet.  Given total $n$ policies to be indexed, the CLP-tree reduces the linear search time $O(n)$ to approximately the height of the tree $O(log(n))$.

The right side of the DLP-tree indexes semantic locations based on the hierarchical relationship among their semantic meanings. In particular, semantic locations are first classified into the basic  categories, such as ``bar", ``hospital", ``shopping mall" and ``company". Basic categories are further classified into more generic categories which are the upper level of the DLP-tree.  For example, basic categories like ``bar" and ``shopping mall" can be classified as a more generic category:  ``entertainment", and basic categories like ``hospital", ``clinic", and ``urgent care" can be classified as  ``medical". Unlike the top-down search in the left side of the DLP-tree, the search in this part of the DLP-tree is from the bottom to the top. This is because users are allowed to specify their sensitive locations using semantic words at different granularity. Some users may specify ``entertainment" in their policies while some users may specify only ``bar" in their policies. Thus, the user policies are attached to different levels of the DLP-tree correspondingly. An entry in a node of this side of the DLP-tree is in the form of $\langle$$\varpi$, $\Xi$, $PPT$$\rangle$, where $\varpi$ is the semantic keyword specified in the list of LAMPi policies (denoted as $\Xi$), and $PPT$ is the pointer to the parent node in the DLP-tree.


The following is an example of how to look up the LAMPi policies that use semantic locations. Given a photo depicting a group of people, assume that its tag indicates it is a bar. In order to check if anyone in this photo considers this place as sensitive, we will search the right side of the DLP-tree. Starting from the leaf level, we find the node that contains the keyword ``bar" and retrieve the associated policy IDs. Then, we visit its parent node with the keyword ``entertainment", and also retrieve all the policies associated with it.  Again, we go up to the parent node with the keyword ``any place",  and retrieve all the policies.

The owners of the policies retrieved from the DLP-tree  will be compared against the people on the photo using face recognition as discussed in the following subsection.

\subsubsection{Speed Up Face  Recognition}

To speed up the individual face comparison, we adopt two strategies. One is to pre-compute the user's face features when the user set up his/her LAMPi policies, which  helps save the face recognition time during the photo uploading phase. The other is to employ  multi-thread programming to conduct individual pairs of face recognition simultaneously.

Specifically, we calculate the face feature using the $load\_photo(image\_path)$ and $encode(image)$ functions in an open source python face recognition tool by Geitgey \cite{facialrec}.  As reported, this face recognition tool has achieve 99.38\% accuracy. The $load\_photo(image\_path)$ function loads an image from an image path using PIL. This image is then converted to a numpy array and returned. Then, the $encode(image)$ function takes the image numpy array as the input, and utilizes a pre-trained algorithm from dlib's facial recognition library to convert the image numpy array to a 128 dimension facial description. We then store the 128 dimension face features along with the user ID for the future face recognition.

Given a new photo, we first detect faces on the photo  using the  $locations(image)$ in the Geitgey's face recognition tool. For the detected faces, we calculate their face features in the similar way as aforementioned. Then, we compare the face features of those in the photo with those associated with the retrieved LAMPi policies that specify the photo location as sensitive. The face comparison is conducted using the $compare(source, destination, tolerance)$ function in the face recognition tool, which takes a source facial feature, a destination facial feature, and a tolerance. The tolerance value is set based on the sensitiveness value in the corresponding LAMPi policy. The function  calculates the Euclidean distance between the facial feature vectors and checks if the  distance is  below the tolerance value. If the distance is smaller than the tolerance value, the two faces are considered match.

It is worth noting that  social network sites can also use their existing face recognition tools when adopting our proposed privacy preservation function.

\subsection{Protect Users with Location Privacy Concerns}

After the face recognition, we will obtain a set of users who are in the photo and concerned about their privacy at the photo location. For these users, we propose to replace their faces  so that they will not be recognizable even if the photo is shared  publicly. There have been several face replacement algorithms and software \cite{face-2d-mapping,facialswap}. We revised an open source software for the face replacement \cite{facialswap} and integrated it into the LAMP system.

Figure~\ref{fig:overview} illustrates a running example of how the LAMP system protects privacy. Assume that a student reporter took some photos on campus and plans to post them online to show the student life at a university in Paris. When she uploaded the photos to the social media site that deployed the LAMP system, the LAMP system will check each photo and conduct the following privacy preservation procedure.

First, the LAMP extracts the photo's metadata to obtain the location information. In the example, the location information includes both the university address and the semantic keyword ``university". Next, the LAMP  system will search the policy database to find the policies which specify the photo's location as sensitive. For example, Bob was conducting an important business at Paris from 11/15/2019 to 12/15/2019, and he would like to keep his photos in Paris private as indicated by his policy $P_i$. Alice, a celebrity,  was taking a year off  to study abroad. Alice does not want to be followed or disturbed by her fans whenever she was at the university, and thus she has set her LAMPi policy as follows: $P_{j}$=$\langle$Loc=Universite Paris Diderot, Typ=E, Int=Anytime,  $\xi$=High$\rangle$. Note that Alice sets the time interval of protection to ``Anytime" for convenience instead of using the exact time and date duration. She can simply remove this policy to release the protection after she returns home.

From the set of retrieved policies, the LAMP system next loads the face feature vectors of these policies' owners. These candidate face features will be used for face recognition, i.e.,  compared with faces on the uploaded photos which are highlighted in boxes in Figure \ref{fig:overview}. In the example, Alice's face was identified (pointed by the red arrow in the figure).   Since Alice has wished to remain private in this location, the LAMP system will then help privatize this user through face replacement.

\begin{figure}[!t]
\begin{subfigure}{3.3in}
\includegraphics[width=3in]{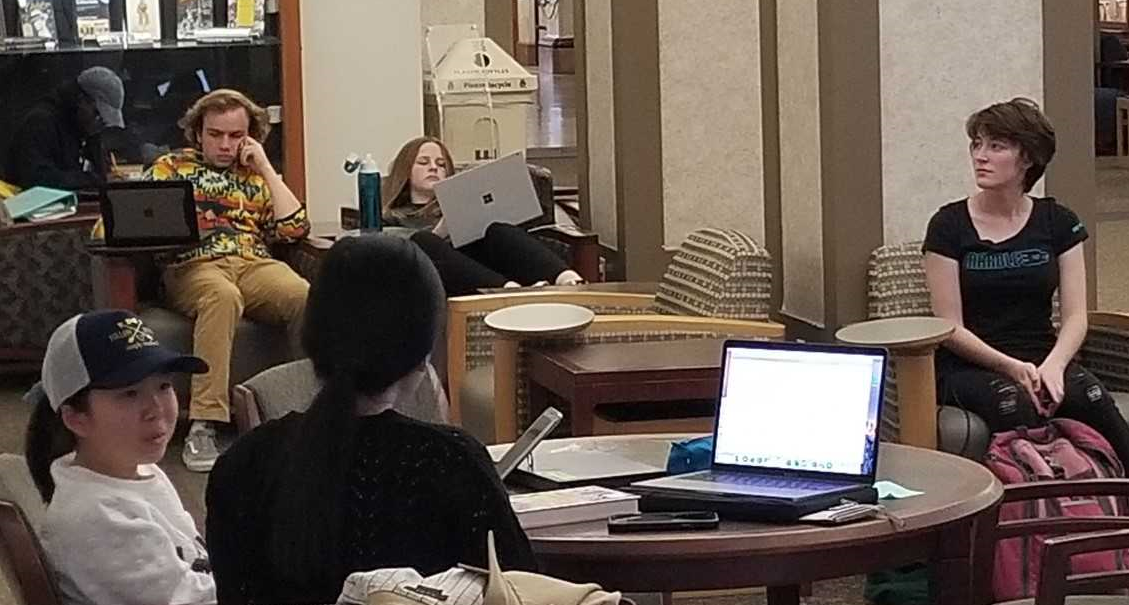}
\caption{Original Photo}
\end{subfigure}
\begin{subfigure}{3.3in}
\includegraphics[width=3in]{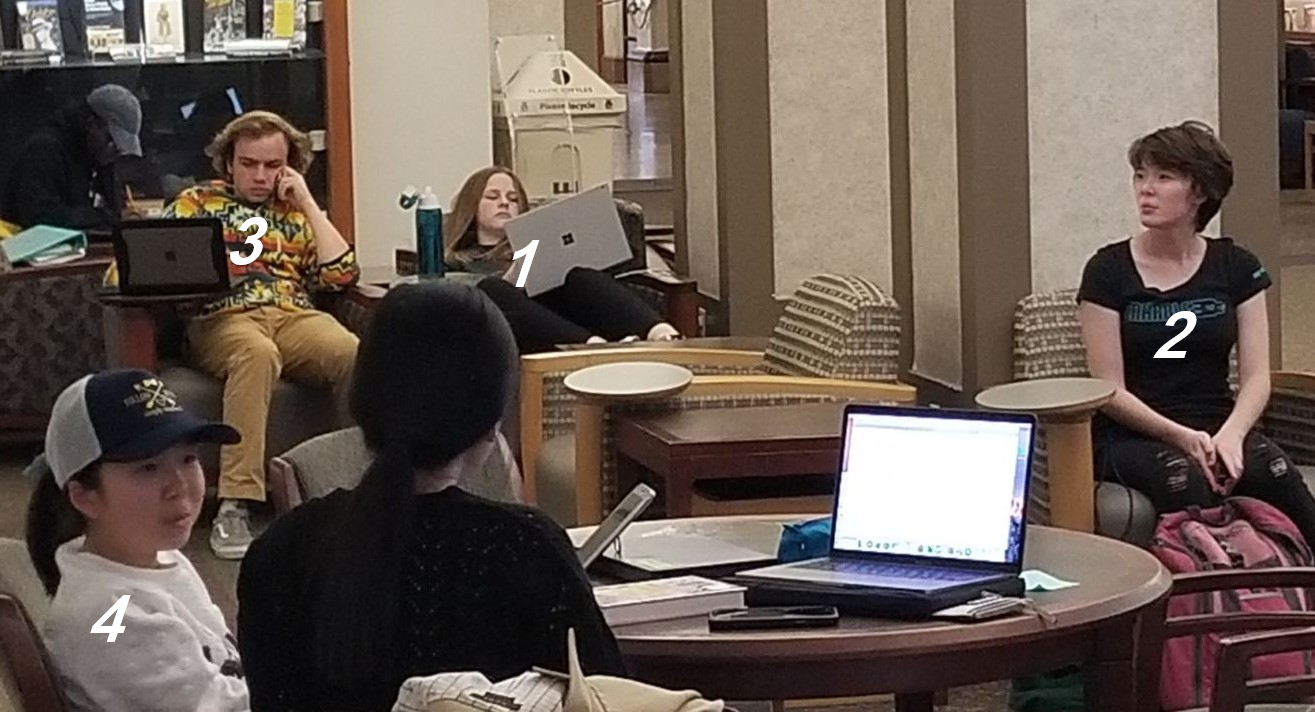}
\caption{Face 2 Replaced}
\end{subfigure}
\caption{Photo Comparison}
\label{fig:swapp}
\end{figure}

In our current implementation of the face replacement, a reference face needs to be manually selected to replace Alice's face. In the example, we chose another female's face who does not have privacy concerns at this location, and use it to replace Alice' face. Figure \ref{fig:swapp}  shows the original photo uploaded by the user and the photo after the face replacement. Specifically, in Figure \ref{fig:swapp} (b), Alice's face (No.2 person) has been replaced with the face of the No. 4 person. Observe that the modified photo looks very natural. Therefore, we expect that the modified face will not raise special attention from users who are viewing the photo.



Future work in this would be to utilize an artificial face generation method so that the face replacement  can be conducted automatically without selecting a candidate face that does not have privacy concerns. Moreover, besides face replacement, it may be interesting to incorporate techniques that  replace other portions of a user such as a user's hair style and attire to prevent someone who is familiar with the user from identifying the person. However, we also argue that since current face replacement results in a natural look, people viewing the photo do not know the photo has been modified or not, and may not try hard to match each person in a photo with someone they know.




\section{Privacy Evaluation}\label{sec:privacy evaluation}

In order to evaluate the effectiveness of our proposed privacy protection, we conducted another round of user study to see if participants are still able to identify the person who requires privacy protection and has been  processed by our system. The idea is to present a set of testing photos and two reference photos to the participants, and ask them to try to identify the female and male references from the photos. Among the testing photos, some contain the referenced female and male without modification which represent the scenarios when people did not mark that location as sensitive; some contain the reference female and male with replaced faces which represent the scenarios when the people require privacy protection at that location; some contain blurred faces which represent the traditional privacy protection approach; and some do not contain any of the referenced people which are used as comparisons. The details of the user study are described as follows.

We have recruited total 102 participants on Mechanical Turk. There are 51 females and 51 males. Among them, 22\% are 18 to 25 years old, 43\% are between 26 and 35, 19\% are between 36 and 45, 12\% are 46 to 55, and 4\% are above 56 years old.  The user study is fully anonymous and follows the IRB exempted project guidelines.



Our study starts by telling participants that they will review images with numbered people within them. They will also have a reference photo for a person's face. They are told that if they can identify the reference person in the photo with a large degree of certainty, they just mark down the number of the person on the photo. They are also told that some photos may not include the reference person, and if they can not identify the reference person with a high degree of certainty, they need to input 0.

Each participant was asked to view 10 images. Each image contains  4 to 10 faces in the foreground and background. Half of these photos were asked about a male reference, and the remaining half were asked about a female reference. Photos for both male and female references included 1 photo with the reference not in the photo, 1 photo with the references whose face have been replaced, 1 photo with the references whose faces have been blurred, and 2 photos of the references in clear view. The photos in clear view give us an understanding of if the majority of participants can correctly identify the person within the image without any changes, and then we can compare that with how they react when they see images that have been modified. Lastly,  we  ask if they noticed any image that has been altered in some way by presenting them unaltered photos and photos with replaced faces.  We  summarize participants' responses using the misidentification ratio which is the percentage of participants who did not correctly identify the reference person in the photo. From the study, we have the following two major findings.

\vspace{5pt}
\noindent {\bf \em Finding 1}: {\em Photos with replaced faces have on average the highest misidentification ratio.} Specifically, as shown in Table \ref{tab:indentifyimages}, 84\% of participants did not recognize the male reference in the photos where his face is replaced. Similarly for the female reference,  77\% of participants did not recognize her in the photo with her face swapped. Both ratios are higher than that for the photos with blurred faces. This is possibly because when a face is blurred in the photo, the participants of the study clearly know which face to examine, and they can pay closer attention to the person of the blurred face including checking the hair style and other features. Thus, more participants were able to guess that the blurred faces were the references with high confidence. When it comes to the photos with swapped faces, the participants do not know which face is swapped. There is more work for them to examine all the  faces in the photo in great details. Thus, fewer people were able to correctly identify the references. In addition, the misidentification ratios for photos with full face shown are much lower than the misidentification ratio of modified faces.

\vspace{5pt}
\noindent {\bf \em Finding 2}:  {\em Unaltered and face swapped photos are hard to distinguish}. At the end of the user study, we present an unaltered photo and a photo that contains a swapped face to the participants. For each photo, the participants were asked to check if the photo has been altered. The results are very interesting.  Given an unaltered photo, 45\% of participants said it had been altered. However, given the photo with a swapped face, only 32\% of participants said it was altered. Such results could be because that participants believed that some photos must be altered since the survey asked the question, so they were taking a guess whether a photo had been altered even though they were not 100\% sure. This interesting result indicates that it would be really hard for human eyes to distinguish a face swapped photo from unaltered photos.

\vspace{5pt}
The result from our study demonstrates the effectiveness of privacy protection of  our proposed  use of face replacement. Moreover, in the real world scenario, when an attacker suspects a blurred face, he can utilize deblurring technique to further verify. In contrast, the replaced faces may not even arouse any attention from viewers, and hence we expect much higher misidentification ratio in the real social media sites.

\begin{table}[!t]
\centering
\small{\caption{Privacy Evaluation Results}
\label{tab:indentifyimages}
\begin{tabular}{|p{5.7cm}|p{2.3cm}|}
\hline
\textbf{Image Type} & \textbf{Misidentification Ratio}  \\
\hline
\hline Male reference with full face shown. & 30\% \\
\hline Female reference with full face shown. & 26\%  \\
\hline Male reference with 40\% of face showing. & 76\%\\
\hline Female reference with 50\% of face showing. & 42\%\\
\hline Male reference not within photo. & 44\%\\
\hline Female reference not within photo. & 11\%\\
\hline Male reference with face blurred. & 79\%\\
\hline Female reference with face blurred. & 68\%\\
\hline Male reference with face swapped. & 84\%\\
\hline Female reference with face swapped. & 77\%\\

\hline
\end{tabular}}
\end{table}

\section{Efficiency Evaluation}\label{sec:efficiencyeval}

In this section, we aim to examine the efficiency and scalability  of our proposed LAMP system. All of our experiments were conducted on custom Intel based computer with an i5-6600k at 3.9 GHz turbo clock, and 24 GB of 2133 Mhz RAM (computer specification).

Users' LAMPi policies are synthetically generated. We randomly select exact locations and semantic locations along with time and date constraints for each user to create the LAMPi policies. In the experiments, we vary the number of sensitive locations (i.e., the number of policies) specified per user to test the efficiency of our algorithms.

We collected 5000 facial images from "Labeled Faces in the Wild" database \cite{LFWTech}. These images are used to simulate uploaded images that we need to check the privacy compliance. Since not all the images that we collected associate with location information and our goal is to evaluate only the efficiency of our algorithms, we randomly generate location tags for each image. Each image is associated with both a coordinate location and 1 to 5 keywords indicating the semantic meanings of the location. The keywords for semantic locations are generated based on a four-level hierarchy similar to the one shown in Figure \ref{fig:dlp}. In the experiments, we vary the total number of distinct  locations to test their impact on our system performance.

\vspace{5pt}
\noindent{\bf (1) Varying the Total Number of Users:} In the first round of experiments, we aim to evaluate how fast our system can retrieve users whose LAMPi policies specify the location of the uploaded photo as sensitive. We vary the total number of users from 100K to 1M. There are total 100K distinct locations. Each location has a unique address and 1 to 5 keywords.  Each location may be marked as sensitive by 1000 users, thus resulting 100M LAMPi policies in total. There are 50\%  policies on exact locations and 50\% on semantic-based locations.

\begin{figure}[!t]
    \centering
    \epsfig{file=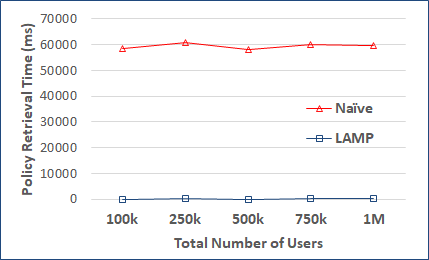,width=0.35\textwidth}
    \caption{Effect of Varying the Total Number of Users}
    \label{exp:user}
\end{figure}

We compare the performance of our system with a naive solution which scans  all the users' policies to identify the match. Figure \ref{exp:user} shows the experimental results. As we can see that, our LAMP system is  thousands of times faster than the naive approach. Specifically, it took the naive algorithm about 1 minute to locate a policy, while the LAMP system needs only 50 milliseconds. This is because our LAMP system indexes policies based on their locations using the DLP-tree.  Considering 100 entries per node in the DLP-tree, 5 levels of the tree will be able to index about 1 billion policies. In other words, given a location either an address or a semantic keyword, we only need to check a few nodes (a few hundred entries out of 1 billion) in the DLP-tree to locate the group of policies that specify this location as sensitive. The naive approach will have to check every policy to see if the policy specifies the photo's location as sensitive, and hence it is extremely slow.

From the figure, we also observe that both approaches are not affected much by the total number of users. This is because the policy retrieval time is  dominated by the total number of policies rather than the total number of users. In this experiment, the total number of policies is 100M regardless of the increase of the users. The differences among the test datasets mainly reside in the policy owners who are selected from a 100K user pool  or a 1M user pool.

\vspace{5pt}
\noindent{\bf (2) Varying the Number of Distinct Locations:} In this round of  experiments, we evaluate the effect of the total number of distinct locations.  We vary the number of distinct locations from 10K to 100K. We set the total number of users to 1M and the number of policies per location to 1000. Under this setting, the total number of policies ranges from 10M to 100M, and each user has 10 to 100 policies, i.e., each user specifies 10 to 100 locations as sensitive.

\begin{figure}[!t]
    \centering
    \epsfig{file=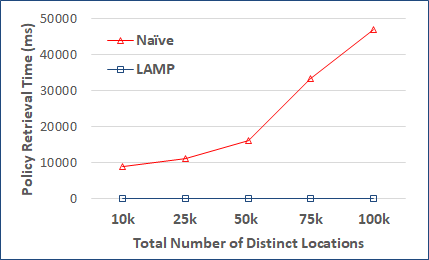,width=0.35\textwidth}
    \caption{Effect of Varying the Total Number of Distinct Locations}
    \label{exp:loc}
\end{figure}

Figure \ref{exp:loc} compares the performance of our approach and the naive approach. Observe that the time taken to retrieve the policies for a given photo by the naive approach increases dramatically with the growth of the number of distinct locations. In contrast, our LAMP approach achieves constant and efficient performance in all the cases. Specifically, the naive approach requires 10 to 50 seconds when the number of locations increases from 10K to 100K, whereas our approach only needs 19 milliseconds to 55 milliseconds. This again demonstrates the effectiveness of the policy organization by the LAMP approach. Unlike the naive approach which needs to check all the policies to find those specifying the photo location as sensitive, the search strategy adopted by our LAMP system significantly reduces the search scope. The size of the DLP-tree is only logarithmic to the  number of distinct locations. Therefore, the processing time of the LAMP system only increases a little.

\vspace{5pt}
\noindent{\bf (3) Varying the Number of Distinct Semantic Keywords:} When testing semantic-based policies, we also set the total number of users to 1M and the total number of exact locations to 100K. Each location is associated with 5 semantic keywords. We vary the total number of distinct keywords from 250 to 5000. 5000 categories of places are considered an extreme case in the real world scenario, and hence we think it is sufficient to be used to test the scalability of our approach.


\begin{figure}[!ht]
    \centering
    \epsfig{file=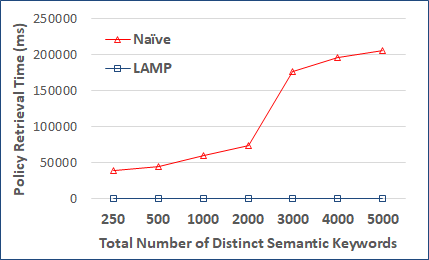,width=0.35\textwidth}
    \caption{Effect of Varying the Total Number of Distinct Semantic Keywords}
    \label{exp:sem}
\end{figure}

Figure \ref{exp:sem} shows the experimental results which demonstrate similar trend as that in the previous experiment regarding the policies on exact locations. Specifically, our LAMP system requires only 49 milliseconds to retrieve policies in the dataset with 250 distinct semantic keywords and 55 milliseconds for the dataset with 5000 distinct keywords. The naive approach took much longer time (more than 200 seconds) especially when the number of distinct keywords increases. This is because the naive approach needs to look through each different semantic keyword until find the matching policies.

\vspace{5pt}
\noindent{\bf (4) Varying the Number of Policies per Exact Locations:} Next, we examine the effect of the number of policies per exact location. We fix the total number of users to be 1M and the total number of distinct locations to be 100K. We vary the number of policies per location from 0.1\% to 1\% of the total number of users. That means the number of policies per location ranges from  1M$\times$0.1\%=1000  to 1M$\times$1\%=10,000.  The total number of policies ranges from 1M to 1 billion. This corresponds to 1 to 1000 policies per user.

\begin{figure}[!ht]
    \centering
    \epsfig{file=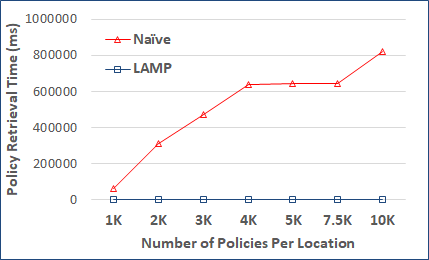,width=0.35\textwidth}
    \caption{Effect of Varying the  Number of Policies per Exact Location}
    \label{exp:polloc}
\end{figure}

The performance comparison between our LAMP system and the naive approach is shown in Figure \ref{exp:polloc}. Observe that the LAMP system significantly outperforms the naive approach. It took similar amount of time for the naive approach regardless the number of policies per location since the naive approach always needs to check all the policies in all locations for a photo. As for the LAMP system, it only checks the policies that contains the photo's location. Thus, when there are more policies per location to be retrieved, the cost of the LAMP system  increases slightly. It is worth noting that our system is still very fast as it needs only 128 milliseconds to find the candidate policies among 1 billion policies.

\begin{figure}[!ht]
    \centering
    \epsfig{file=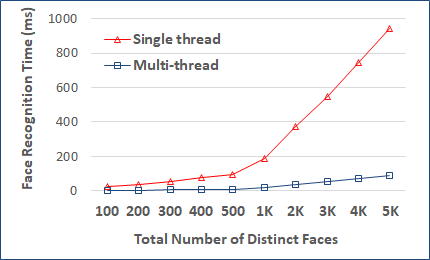, width=0.35\textwidth}
    \caption{Face Recognition Time (18 faces in the uploaded photo) }
    \label{exp:facerec}
\end{figure}

\vspace{5pt}
\noindent{\bf (5) Performance of Face Recognition and Replacement:} After locating candidate users who specify the photo's location as sensitive, the next step is to check if the user actually appears on the photo through face recognition.  Note that in our system, we only need to compare faces in the photo with candidate users who specify the photo's location as sensitive, rather than the users in the whole social network. Therefore, we vary the number of candidate users from 100 to 5000 whereby the value 5000  conforms with our previous setting of 5000 policies per location. We select a group photo that contains 18 faces as the photo to be uploaded. That means, there will be up to 5000*18=90,000 face comparisons. We tested both linear and  parallel face recognition performance. As shown in Figure \ref{exp:facerec}, the linear algorithm starts to struggle with the increase of the number of candidate users. Our parallel algorithm performs relatively constantly and the face recognition time stays below 100 milliseconds in all cases. We expect the face recognition to be even faster at the real server which has better hardware and can spawn more concurrent threads.

We also check how long it takes to replace a user's face. Since face replacement is not the research focus of our work, we adopt an existing open source software \cite{facialswap} to gain the idea of the time needed for face swapping. We observe that it took about 8 seconds to replace a face. We expect the face replacement to be faster at the server side, and a few seconds of delay before the photo go live online would not be very noticeable by users considering there are also network and webpage refresh delays.

\section{Related Work}\label{sec:related}

\vspace{5pt}
\noindent {\bf (1) Image Privacy Protection for the Photo Owner}

There have been a large body of image privacy protection works which focus on protecting the privacy of the photo owners \cite{context-dependent-ml, a3p, privacy-level,user-defined-sharing,cenlocshare}. These types of solutions include frameworks meant for suggesting privacy policies (i.e., which groups of people to share the image) for the photo  owners  at the time of the image being uploaded.  For example, an early work \cite{a3p} is by Squicciarini et al. who propose a privacy policy prediction system called A3P which considers  image content, image metadata as well as the photo owners' historic privacy preferences when generating the policies.  In \cite{privacy-level}, Hu et al. propose an interesting idea of calculating  a level of sensitivity for each photo based on both  user-defined  rules and general rules discovered by machine learning. Users can then use the sensitivity levels as guidance for their privacy settings.  In  \cite{context-dependent-ml}, Yuan et al. employ machine learning algorithms to analyze a social media user's photo sharing behavior, taking into account both the content of the image and the social context of the users who may see the photo.  From that information, the system then determines whether or not to share the photo, entirely or partially, with a certain user.

\vspace{5pt}
\noindent {\bf (2) Multi-Party Image Privacy Protection}

The protection for a single photo owner has later been  extended to co-owners of the photo, i.e., people who took a group photo together. This type of multi-party privacy protection is typically achieved by considering privacy preferences of each party,   solve policy conflicts among multiple parties, and then blur the faces with privacy restriction  \cite{sharing-google+,hideme,iprivacy,multiparty-authorization-2011,multiparty-access-2013,face-off}. For example, Hu et al. \cite{multiparty-access-2013} define an access control model to capture the multiparty authorization requirements, based on which they develop  multiparty policy specification scheme and algorithms to solve policy conflicts  among multiple parties. llia et al.  \cite{face-off}   propose a new way for multi-party privacy protection. They employ face recognition to automatically detect faces on the photo,  and present  the  photo  with  the  restricted  faces  blurred  out.  Similarly, Fan et al. \cite{iprivacy} also leverage the blurring technique to enforce privacy policies. They  cluster semantically similar images and generate common privacy settings for privacy-sensitive classes. Then, the proposed iPrivacy system  automatically identifies privacy sensitive objects on images and also help blur the detected sensitive objects.

Unfortunately, very limited efforts have been devoted into privacy protection of people who occur in the background of others' photos like what we discuss in our work. As acknowledged in \cite{face-off},  identifying a person who is not related to the photo owner, i.e., not in the photo owner's contact list, would require a huge amount of computing resources due to the need to  scan the whole enormous social network user set. A related work by Henne et al. \cite{comment6} needed up to 1 hour to check an image and notify the bystanders.  It detects only 50\%-70\% privacy violations in many cases and did not enforce the protection.



\vspace{5pt}
\noindent {\bf (3) Privacy Issues Regarding Photo Metadata}

Besides achieving protection through proper policy configurations, recent studies also look into potential privacy breach caused by metadata associated with photos \cite{digital-metadata}. Metadata like geotags and timestamps can easily disclose a person's location information, and multiple photos with geo-tags and timestamps may be used to track a person. To prevent undesired exposure, researchers \cite{hidden-risks} have proposed  to remove metadata. However, such strategy may not be sufficient since the context of the photo may still reveal the location with the advance of the image processing technology. Our approach takes another route by hiding the person's face so as to avoid any location privacy breach. In another  work \cite{ripa},  Chandra et al. developed a mobile app  which can detect human subjects  and issue a privacy alert if the  location is sensitive. Their location privacy protection component is relatively preliminary which simply stores users'  sensitive locations  as link lists.

\vspace{5pt}
\noindent {\bf The Uniqueness of Our Work}

To sum up, our work is different from existing works in terms of the following aspects. First, we aim to protect privacy of every human subject depicted on a photo regardless he/she is the owner of the photo or happens to appear in the background of the photo. Second, we aim to preserve location privacy during the photo sharing. We not only propose a sophisticated location-aware image privacy  policy language, but also design an efficient and scalable approach that minimizes computational overhead incurred  by the privacy protection process. Third,  we go one step further by providing  privacy protection through face replacement instead of only preventing photos from being shared. Face replacement would also provide a better protection than face blurring since images with blurry faces may invite curiosity and become  targets.


\section{Conclusion}\label{sec:conclusion}

In this paper, we propose a novel idea to address the increasing concerns of location tracking of an individual  through online images posted by others. Specifically, we define a new access control model, namely Location-Aware Multi-Party image (LAMPi) access control, which goes beyond the traditional access control that offers protection to only the owners of the image. Our proposed LAMPi access control mechanism provides equal privacy protection to  every human subject on an online photo, no matter the human subject is the owner of the image or not, is  at the foreground or background of the image. We also design an efficient policy management system that leverages policy indexing techniques and uses face replacement as policy enforcement, and we achieve the privacy protection in real time of photo uploading process. Our user studies and experimental results on the system prototype demonstrate both effectiveness and efficiency of our approach.






\bibliographystyle{plain}
\bibliography{ref}


\end{document}